\documentclass{article}
\pdfoutput=1

\usepackage{listings}

\usepackage{xcolor}

\usepackage{graphicx}

\usepackage{xspace}

\usepackage{todonotes}

\usepackage{tikz}
\usepackage{changepage}

\usepackage{amsmath,amssymb}

\newcommand*\circled[2][black]{\tikz[baseline=(char.base)]{
            \node[shape=circle,fill,inner sep=2pt,fill=#1] (char) {\textcolor{white}{#2}};}}
\newcommand*\squared[2][black]{\tikz[baseline=(char.base)]{
              \node[shape=rectangle,fill,inner sep=2pt,fill=#1] (char) {\textcolor{white}{#2}};}}
\makeatletter
\DeclareRobustCommand\onedot{\futurelet\@let@token\@onedot}
\def\@onedot{\ifx\@let@token.\else.\null\fi\xspace}

\def\eg{\emph{e.g}\onedot} 
\def\ie{\emph{i.e}\onedot} 
 
\def\etc{\emph{etc}\onedot}

\makeatother

\graphicspath{ {./figures/} }

\definecolor{codegreen}{rgb}{0,0.6,0}
\definecolor{codegray}{rgb}{0.5,0.5,0.5}
\definecolor{codepurple}{rgb}{0.58,0,0.82}
\definecolor{backcolour}{rgb}{0.95,0.95,0.92}
\definecolor{deepblue}{rgb}{0,0,0.5}
\definecolor{deepred}{rgb}{0.6,0,0}
\definecolor{deepgreen}{rgb}{0,0.5,0}

\lstdefinestyle{mystyle2}{
  language=Python, 
  morekeywords={search,rule,bm25_match,neural_match,neural_extract,op_filter,bm,_match,translate,answer,eval,fact},
  otherkeywords={0,1,2,3,4,5,6,7,8,9,=,<,>,ids(),vars()},
  backgroundcolor=\color{white},%
  commentstyle=\color{codegreen},
  keywordstyle=\color{codepurple},%
  numberstyle=\tiny\color{codegray},
  stringstyle=\color{deepred},%
  basicstyle=\ttfamily\normalsize,%
  emph={_asin , _total_reviews , _price , _title , _review , _review_text , _answers, NeuroQL},     
  emphstyle=\color{blue}, 
  breakatwhitespace=false,         
  captionpos=b,                    
  keepspaces=true,                 
  numbers=left,                    
  numbersep=5pt,                  
  showspaces=false,                
  showstringspaces=false,
  showtabs=false,                  
  tabsize=2,
}

\lstdefinestyle{datastyle}{
  language=Python, 
  backgroundcolor=\color{white},%
  commentstyle=\color{codegreen},
  keywordstyle=\color{codepurple},%
  otherkeywords={title,price,brand,stars,total_,total_reviews,review,item_weight,item_weight_units,item_model_number,is_discontinued_by_manufacturer,?asin,?title,?price,ids,search,==,translate,answer,answers,eval,fact},
  numberstyle=\tiny\color{codegray}, 
  stringstyle=\color{deepred},%
  basicstyle=\ttfamily\normalsize,%
  emph={B00001P4ZH, B000AJIF4E},     
  breakatwhitespace=false,         
  captionpos=b,                    
  keepspaces=true,                 
  numbers=left,                    
  numbersep=5pt,                  
  showspaces=false,                
  showstringspaces=false,
  showtabs=false,                  
  tabsize=2
}

\PassOptionsToPackage{numbers, compress}{natbib}

\usepackage[preprint]{neurips_2020}
\usepackage[utf8]{inputenc} %
\usepackage[T1]{fontenc}    %
\usepackage[colorlinks=true,allcolors=blue]{hyperref}       %
\usepackage{url}            %
\usepackage{booktabs}       %
\usepackage{amsfonts}       %
\usepackage{nicefrac}       %
\usepackage{microtype}      %
\usepackage{enumitem}

\usepackage{amsfonts}%
\usepackage{algorithmic}
\usepackage{graphicx}
\usepackage{textcomp} 
\usepackage{dutchcal}
\usepackage{nccmath}
\usepackage{stmaryrd}
\usepackage{tabu}
\usepackage{longtable}

\makeatletter

\newcommand{\leqnomode}{\tagsleft@true}
\newcommand{\reqnomode}{\tagsleft@false}

\makeatother

\definecolor{tableHeader}{RGB}{200,200,200}%
\definecolor{tableLineOne}{RGB}{245, 245, 245} 
\definecolor{tableLineTwo}{RGB}{224, 224, 224}

\usepackage{xcolor}
\usepackage{mathtools}

\makeatletter
\newcommand{\redub}{}
\def\redub#1{%
  \@ifnextchar_%
    {\@redub{#1}}
    {\@latex@warning{Missing argument for \string\redub}\@redub{#1}_{}}%
}
\def\@redub#1_#2{%
    \colorlet{currentcolor}{.}%
    \color{blue}%
    \underbrace{\color{currentcolor}#1}_{\LARGE\color{black}#2}%
    \color{currentcolor}%
}
\makeatother

\usepackage{tikz}
\usetikzlibrary{arrows,calc,shapes}

\usepackage{balance}
\usepackage{xspace}
\usepackage{hyperref}
\usepackage{amsmath}
\usepackage[most]{tcolorbox}
\usepackage{tabularx}
\definecolor{tableShade}{gray}{0.9}

\tcbset{colback=white,colframe=white,coltitle=black,%
        highlight math style= {enhanced, %
            ,boxsep=0pt}
        }

\usepackage{booktabs}   %
                        
\usepackage{xcolor}
\usepackage{todonotes}
\usepackage{rotating}

\usepackage{subcaption} %
\usepackage[utf8]{inputenx} %
\usepackage{csquotes}       %
\usepackage{microtype}      %
\usepackage{mathtools}      %
\usepackage{bussproofs}
\usepackage{multicol}
\usepackage{etoolbox}
\usepackage{listings}
\newtoggle{InString}{}%
\togglefalse{InString}%

\newcommand*{\ColorIfNotInString}[1]{\iftoggle{InString}{#1}{\color{red}#1}}%
\newcommand*{\ProcessQuote}[1]{#1\iftoggle{InString}{\global\togglefalse{InString}}{\global\toggletrue{InString}}}%
\definecolor{OliveGreen}{rgb}{0,0.6,0}

\makeatletter
\lstdefinestyle{mystyle}{
  basicstyle=%
    \ttfamily
    \lst@ifdisplaystyle\footnotesize\fi
}
\makeatother

\usepackage{pifont}

\usepackage{tikz}
\usetikzlibrary{arrows,shapes,automata,petri,positioning}
\tikzset{
  place/.style={
      circle,
      thick,
      draw=blue!75,
      fill=blue!20,
      minimum size=6mm,
  },
  transitionH/.style={
      rectangle,
      thick,
      fill=black,
      minimum width=8mm,
      inner ysep=2pt
  },
  transitionV/.style={
      rectangle,
      thick,
      fill=black,
      minimum height=8mm,
      inner xsep=2pt
  }
}

\usepackage{lipsum}
\usepackage[absolute, overlay]{textpos} %

\usepackage[nounderscore]{syntax}
\usepackage{enumitem}
\newcounter{descriptcount}

\usepackage{xspace}

\usepackage{tcolorbox}
\usepackage{float}

\title{DeepLL: Considering Linear Logic for the Analysis of Deep Learning Experiments}

\newcommand{\authorbreak}{\protect\\[-16pt]}
\author{%
  Nick Papoulias\authorbreak
  Director of Research\authorbreak
  OrgD. Labs\authorbreak
  \url{https://orgdlabs.com}\authorbreak
  \texttt{npapoylias@orgdlabs.com}
}

\begin{document}

\maketitle

\begin{abstract}
Deep Learning experiments have critical requirements regarding the careful handling of their datasets as well as the efficient and correct usage of APIs that interact with hardware accelerators. On the one hand, software mistakes during data handling can contaminate experiments and lead to incorrect results. On the other hand, poorly coded APIs that interact with the hardware can lead to sub-optimal usage and untrustworthy conclusions. In this work we investigate the use of Linear Logic for the analysis of Deep Learning experiments. We show that primitives and operators of Linear Logic can be used to express: \textit{(i)} an abstract representation
of the control flow of an experiment, \textit{(ii)} a set of available experimental resources, such as API calls to the underlying data-structures and hardware as well as \textit{(iii)} reasoning rules about the correct consumption of resources during experiments. Our proposed model is not only lightweight but also easy to comprehend having both a symbolic and a visual component. Finally, its artifacts are themselves proofs in Linear Logic that can be readily verified by off-the-shelf reasoners.    
\end{abstract}
 
\section{Introduction}\label{sec:intro}

\subsection*{Deep Learning Requirements}

The increased usage of Deep Learning \cite{lecun2015deep,goodfellow2016deep,bishop2023deep} experiments in all areas of science
and engineering, necessitates the careful study of current experimental practices. This is a pressing concern, especially because most practitioners are domain experts in their respective fields and not software engineers or computer scientists. Moreover, the increased usage of generated code in experiments can introduce problems that are harder to detect. 

In this work we observe that experimental settings \cite{chollet2024deep, howard2020deep} in Deep Learning, have both: 

\begin{enumerate}[label=\textbf{\textit{\alph*}.}]
\item \textbf{\textit{Critical data-provenance requirements}} that mandate the careful handling of datasets during different experimental phases (such as \textit{training}, \textit{validation} and \textit{testing}) and
\item \textbf{\textit{Strong correctness and efficiency requirements}} for the usage of modern APIs that interact with hardware accelerators (such as GPUs and TPUs \cite{wang2019benchmarking})
\end{enumerate}

On the one hand, software mistakes during data handling can contaminate experiments and lead to incorrect results. Indeed if data-structures (such as lists, sets, tensors, etc) that are meant to be used in different parts of the experiments are not handled correctly, they can invalidate all metrics relating to accuracy and generalization of a model. Consider for example the crucial separation between training and validation datasets. If data from the validation set is used for training or if data from the training set are leaked into the validation set, then \textit{(i)} all metrics regarding overfitting during training are invalidated and consequently \textit{(ii)}  all metrics and results considering the ability of a model to generalize are also put into question. Crucially, since the validation set itself is used to tune hyperparameters (such as learning rate, batch size, etc) during training, the existence of yet a third set of data (the test set) is required to confirm the generalization of a model in completely unseen data after training \cite{stevens2020deep}.

\begin{figure*}[!t]
  \advance\leftskip-2cm 

  \begin{subfigure}{\textwidth}
      \centering
      \begin{tcolorbox}[text width=16.5cm]%
          \centering
          \circled[black!30!red]{$e_{}$} Experiment\hspace{2em} \circled[black!30!orange]{$t_{}$} Training\hspace{2em} \circled[black!30!yellow]{$f_n$} Function in path\hspace{2em} \circled[black!30!green]{$m_{}$} API as resource\hspace{2em} \circled{$\pi_n$} Path transition\hspace{2em} 
      \end{tcolorbox}
  \end{subfigure}
  \vspace{0.5em}

\begin{subfigure}{.7\textwidth}

\begin{lstlisting}[language=Python]

  #Experimental Environment of Interest /*!\circled[black!30!red]{$e_{}$}!*/ 
  training_slice = None
  validation_slice = None
  testing_slice = None
  model = None

  #Data Processing
  def load_training_slice(dataset):
    ...

  def load_validation_slice(dataset):
    ...

  def load_testing_slice(dataset):
    ...

  #Model Design & API
  def create_model():
    ...

  def forward_pass(model, x): /*!\circled[black!30!yellow]{$f_1$}!*/
    return model(x) /*!\circled[black!30!green]{$m_{}$}!*/
  
  def forward_sampling_pass(model, x, t): /*!\circled[black!30!yellow]{$f_2$}!*/
    return model(x, t).sample /*!\circled[black!30!green]{$m_{}$}!*/

  #Experimental Phases
  def training(): /*!\circled[black!30!orange]{$t_{}$}!*/
    ...
    if needs_sampling:
      model_output = \
        forward_sampling_pass(model, x, t)
    else:
      model_output = forward_pass(model, x)
    ...

  def validation():
    ...

  def testing():
    ...

\end{lstlisting}

\label{fig:subA}
\end{subfigure}%
\begin{subfigure}{.6\textwidth}
\centering
\begin{tcolorbox}[text width=7.4cm,text height=7.0cm,%
  arc=0mm,title=\ \ \ \ \ \ \ \ Control-flow and resources as a petri net]
\vspace{0.7em}
\begin{tikzpicture}[node distance=0.5cm and 1cm,>=stealth',bend angle=45,auto]
  \node [place,tokens=1,label=above:$e$,fill=black!30!red] (m) {};
  \node [transitionV,label=above:$\pi_1$] (t1) [right=of m] {} edge[pre] (m);
  \node [place,label=above:$t$,fill=black!30!orange] (w) [right=of t1] {} edge[pre] (t1);
  \node [transitionV,label=above:$\pi_{2a}$] (t2) [above right=of w] {} edge[pre] (w);
  \node [transitionV,label=above:$\pi_{2b}$] (t3) [below right=of w] {} edge[pre] (w);
  \node [place,label=above:$f_1$,fill=black!30!yellow] (e1) [right=of t2] {} edge[pre] (t2);
  \node [transitionV,label=above:$\pi_3$] (t4) [above right=of e1] {} edge[pre] (e1) edge[post,out=110,in=50,looseness=1,overlay] (m);
  \node [place,label=above:$f_2$,fill=black!30!yellow] (e2) [right=of t3] {} edge[pre] (t3);
  \node [transitionV,label=above:$\pi_4$] (t5) [below right=of e2] {} edge[pre] (e2) edge[post,out=-110,in=-50,looseness=1,overlay] (m);
  \node [place,tokens=1,label=above:$m$,fill=black!30!green] (l) [below right=of e1] {} edge[post] (t2) edge[post] (t3);
\end{tikzpicture}
\end{tcolorbox}

    \label{fig:subB}

      \centering
      \begin{tcolorbox}[text width=6.5cm,%
          arc=4mm,title=\ \ \ \ Control-flow and resources in Linear Logic]
              \begin{itemize}\large
                  \item[--] $M = e,m$
                  \item[--] \squared{$\pi_1$} = $!(e \multimap t)$
                  \item[--] \squared{$\pi_2$} = $!(t \otimes m \multimap f_1 \& f_2)$
                  \item[--] \squared{$\pi_3$} = $!(f_1 \multimap e)$
                  \item[--] \squared{$\pi_4$} = $!(f_2 \multimap e)$
                  \item[--] $\Pi$ = $\pi_1, \pi_2, \pi_3, \pi_4$
              \end{itemize}
      \end{tcolorbox}
      \small\colorbox{tableHeader}{\textbf{\normalsize Does $\Pi,M \vdash e$ hold for all paths and models ?}}
      \label{fig:subC}  
  \end{subfigure}
  \caption{A first approximate mapping of a training phase (\textbf{left}), with its execution paths and resources (petri net, \textbf{top-right}), mapped into propositional Linear Logic (\textbf{bottom-right}).}
  \label{fig:RunningEg}
\end{figure*}

Keeping track of these different sets is a non-trivial task, since they are frequently mere slices of the same dataset. Moreover information leakage between these sets can take subtler forms other than erroneous usage of a train sample during validation. For example if augmentation \cite{howard2020deep,chollet2024deep} operations to enrich the dataset is performed before slicing, then we may not have an exact sample leaked but a variation of it. This variation still contains information that can invalidate our training. Resources wasted from these mistakes are not mere computational but also economic, since a lot of these experiments have runtimes measured in months that require expensive hardware.

On the other hand, poorly coded APIs that interact with accelerator hardware can lead to sub-optimal usage (with significant economic cost) but also to incorrect conclusions, if there is no way to verify their correctness. These problems become even more pressing with the increased usage of generated code in experiments (such as code generated by Large Language Models \cite{alammar2024hands, sanseviero2024hands}) that can suffer from subtle bugs, version mismatches and hallucinations \cite{salvagno2023artificial} that are harder to detect. 

\subsection*{Linear Logic Reasoning}

\begin{figure*}[h!t]
\begin{tcolorbox}[width=\textwidth,colbacktitle=tableHeader,colback={white},title={\textbf{Inference rules in sequent calculus}},outer arc=0mm,colupper=black]        
   The inference rules for left (l) or right (r) introduction of \\multiplicative conjunction (resource-and) $\otimes$, can be expressed as follows:
  
  \begin{multicols}{2}
  
  \begin{itemize}
  
  \item \begin{prooftree}
  \AxiomC{$M,A,B \vdash \gamma$}
  \LeftLabel{$\Pi$}
  \RightLabel{$\ \ \otimes l$}
  \UnaryInfC{$M,A \otimes B \vdash \gamma$}
  \end{prooftree}
  
  \end{itemize}
  
  \begin{itemize}
  \item \begin{prooftree}
  \AxiomC{$M_1 \vdash A$}
  \AxiomC{$M_2 \vdash B$}
  \LeftLabel{$\Pi$}
  \RightLabel{$\ \ \otimes r$}
  \BinaryInfC{$M_1,M_2 \vdash A \otimes B$}
  \end{prooftree}
  \end{itemize}
  \end{multicols}
  \vspace{0.5cm}
  
  The inference rules for introducing linear implication (lolli) $\multimap$:
  
  \begin{itemize}\vspace{0.5em}
  \item \begin{prooftree}
  \AxiomC{$M_1 \vdash A$}
  \AxiomC{$M_2,B \vdash \gamma$}
  \LeftLabel{$\Pi$}
  \RightLabel{$\ \ \multimap l$}
  \BinaryInfC{$M_1,M_2,A \multimap B \vdash \gamma$}
  \end{prooftree}
  \end{itemize}
  
  \vspace{0.3cm}
  The inference rules for additive conjunction (choice / fork) $\&$, can be expressed as: %
  
  \begin{multicols}{2}
  \begin{itemize}
      \item \begin{prooftree}
      \AxiomC{$M,B \vdash \gamma$}
      \LeftLabel{$\Pi$}
      \RightLabel{$\ \ \& l_A$}
      \UnaryInfC{$M,A \&B \vdash \gamma$}
      \end{prooftree}
  \end{itemize}
  
  \begin{itemize}
      \item \begin{prooftree}
      \AxiomC{$M,A \vdash \gamma$}
      \LeftLabel{$\Pi$}
      \RightLabel{$\ \ \& l_B$}
      \UnaryInfC{$M,A \&B \vdash \gamma$}
      \end{prooftree}
  \end{itemize}

  \end{multicols}
  
  \vspace{0.3em}
  Finally, the equivalent of an identity axiom can be expressed as:
  
  \begin{itemize}\vspace{0.5em}
  \item \begin{prooftree}
  \AxiomC{$$}
  \RightLabel{$id$}
  \LeftLabel{$\Pi$}
  \UnaryInfC{$\gamma \vdash \gamma$}
  \end{prooftree}
  \end{itemize}
  \caption{The sequent calculus, describing the inference rules for our linear logic model.}
  \label{fig:seqCalculus}
  \end{tcolorbox}
  \end{figure*}

In this work, in order to mitigate these issues, we investigate the use of Linear Logic for the analysis of Deep Learning experiments. We show that primitives and operators of Linear Logic \cite{girard1995linear} can be used to express: \textit{(i)} an abstract representation
of the control flow of an experiment, \textit{(ii)} a set of available experimental resources, such as API calls to the underlying data-structures and hardware as well as \textit{(iii)} reasoning rules about the correct consumption of resources during experiments. Equipped with these three components, we can use Linear Logic to statically validate desired properties of experiments. Compared to other methods of static analysis, our proposed model is not only lightweight but it is also easier to comprehend having both a symbolic and a visual component. 

Contrary to classical or intuitionistic logic, Linear Logic explicitly models 
resources by enabling the creation and consumption of ephemeral facts \cite{di2006linear,marti1989petri}. This application of Linear Logic enables us to precisely express and verify the expected usage pattern of API calls for each experimental phase of interest. Moreover it allows us to automatically verify the correctness of our desired properties, by producing as output Linear Logic proofs that can be verified by off-the-shelf reasoners (such as Celf \cite{schack2008celf}). This fact in turn limits the size of the software base that needs to be trusted for our verification process.

Our analysis abstracts the execution of a program in the form of linear logic expressions that are automatically derived from the program's source code. These linear logic expressions can themselves be rendered executable in the form of linear logic programs. Finally, the execution of the derived linear logic programs -- not a separate post-processing step -- produces a proof (or disproof) of the experimental properties under investigation. 

\section{Analyzing Experiments with Linear Logic}
\label{sec:background}

In this Section we present the main intuition behind our verification approach using a simplified example. We then provide our readers with enough background on Linear Logic to understand the underlying structure of proofs involving experimental properties in linear logic.

Figure \ref{fig:RunningEg} shows a common template for Deep Learning experiments involving the training, validation and testing of neural networks. We can distinguish the following parts in this and similar settings:

\begin{itemize}
    \item An \lstinline[language=Python]!#Experimental Environment! section, shown in lines 2 to 6 of Figure \ref{fig:RunningEg}, which define the main experimental entities of interest. These entities form a global state for the experiment. In this simple example, we have included the different dataset slices, such as the \lstinline[language=Python]!training_slice, validation_slice testing_slice! as well as the \lstinline[language=Python]!model! itself. These are mere definitions at this point and can take many forms (global variables, class or instance members of dedicated classes \etc). Without loss of generality we have restricted ourselves to entities that we discuss as examples in this paper. For reference, the experimental environment typically includes numerous and more involved entities such as \lstinline[language=Python]!devices, loss_functions, optimizers, schedulers, metrics! \etc
    \item A \lstinline[language=Python]!#Data Processing! section, as shown on lines 8 to 16 of Figure \ref{fig:RunningEg}, which typically includes the pre-processing steps and slicing of a dataset into the training, validation and testing sets. Here we show example signatures of slicing functions, such \lstinline[language=Python]!load_training_slice! which given a dataset, will pre-process and load training samples into the experimental environment. Considering here that in the majority of experimental setups all slices ultimately come from the same dataset, it is imperative to be able to reason about data isolation and potential information leaks between these sets.
    \item A \lstinline[language=Python]!#Model Design & API! section, as shown on lines 18 to 26 of Figure \ref{fig:RunningEg}. This section is responsible for defining the architecture of model, and a series of helper API calls for dealing with said model. In our example we show examples of simple forward passes through models, with two different calls \lstinline[language=Python]!forward_pass! and \lstinline[language=Python]!forward_sampling_pass!.
    \item An \lstinline[language=Python]!#Experimental Phases! section, as shown on lines 29 to 41 of Figure \ref{fig:RunningEg}. This section contains logic for the different phases of the experiment, such as the \lstinline[language=Python]!training!, \lstinline[language=Python]!validation! and \lstinline[language=Python]!testing! phases. In our example we show a small snippet that is part of the training phase (on lines 31 to 35), that depending on whether the model architecture \lstinline[language=Python]!needs_sampling! or not, will call a different API helper function.
\end{itemize}

\subsection{Analysis with propositional Linear Logic}

In the upper right part of Figure \ref{fig:RunningEg} we see the possible execution paths of an invocation the \lstinline[language=Python]!training! phase (starting on line 29) of our experiment. These possible execution paths starting from \lstinline[language=Python]!training! form a directed graph, which we model as a petri net \cite{peterson1977petri,reisig2013understanding}, \ie a directed graph consisting of (a) places (\eg the execution points $e,t,f_1,f_2$ and $m$ in Figure \ref{fig:RunningEg}), (b) transitions (\eg $\pi_1,\pi_2,\pi_3$ and $\pi_4$) between the aforementioned execution points and (c) tokens (\eg the black marks seen in places $e$ and $m$ of Figure \ref{fig:RunningEg}). The choice of petri nets as a graphical representation of the execution graph is not coincidental. Our goal in this first example is to show how both resources and execution flow can be modeled through Linear Logic (\ie the substructural logic introduced by Jean-Yves Girard \cite{girard1987linear}), seen in the lower right part of Figure \ref{fig:RunningEg}. Petri nets, have been formally shown to accurately depict (in a graphical way) simple propositional Linear Logic models \cite{marti1989petri,brown1995categorical}. These graphical depictions can indeed help practitioners build a basic intuition while working with Linear Logic models.

The execution graph of the \lstinline[language=Python]!training! phase proceeds as follows: The environment \circled[black!30!red]{$e_{}$} first invokes the \lstinline[language=Python]!training! method \circled[black!30!orange]{$t_{}$}, with the current execution point represented by the token starting at \circled[black!30!red]{$e_{}$}. This input token will subsequently cause the firing of the petri net transition \squared{$\pi_1$}. This transition has an equivalent expression in propositional Linear Logic (in the bottom right part of Figure \ref{fig:RunningEg}). We can translate one to the other, if we consider petri net transitions as named logical implications and petri net places as linear logic propositions. For example in the case of transition \squared{$\pi_1$}, we get the equivalent logical implication $\pi_1$, where the places \circled[black!30!red]{$e_{}$} and \circled[black!30!orange]{$t_{}$} become the linear propositions $e$ and $t$. The firing of transition \squared{$\pi_1$}, consuming the input token in \circled[black!30!red]{$e_{}$} and producing a new token in \circled[black!30!orange]{$t_{}$}, can be represented as a linear implication ($\multimap$) between the propositions $e$ and $t$, as follows: $e \multimap t$. Finally, since this implication is part of a static petri net structure (\ie that is not consumed), we annotate it with the linear bang ($!$) operator, signaling that it is a permanent resource which can be re-used: $!(e \multimap t)$.

Subsequently, from \circled[black!30!orange]{$t_{}$} (for the petri net transition \squared{$\pi_2$}) there are two distinct paths that can be taken (depending on the value of the boolean \lstinline[language=Python]!needs_sampling!), leading to either the \circled[black!30!yellow]{$f_1$} or \circled[black!30!yellow]{$f_2$} states. This transition \squared{$\pi_2$} is predicated on the availability of the \circled[black!30!green]{$m_{}$} token, which in this case models the availability of the forward pass calls \lstinline[language=Python]!model(x)! and \lstinline[language=Python]!model(x,t)!, respectively. Both alternatives (i.e $\pi_{2a}$ and $\pi_{2b}$) can be modeled in Linear Logic as a single formula, thusly: $!(t \otimes m \multimap f_1 \& f_2)$. Here the linear multiplicative conjunction connective ($\otimes$) is used, to group the input tokens together in the left part of the implication. Whereas, in the right part of the implication the additive conjunction connective ($\&$) is used, to model the two alternative outputs $f_1$ and $f_2$.
From there and only if the needed resources are available and properly consumed, are we able to successfully return back to the environment \circled[black!30!red]{$e_{}$}, through the \squared{$\pi_3$} and \squared{$\pi_4$} transitions, expressed in Linear Logic as $!(f_1 \multimap e)$ and $!(f_2 \multimap e)$ respectively. Finally, if we define the initial state of the tokens for our petri net as $M=e,m$ and our transitions as the set of linear logic implications $\Pi=\pi_1,\pi_2,\pi_3,\pi_4$, then the sequent: \colorbox{tableHeader}{$\Pi,M \vdash e$}, asks if the training phase can successfully return (\ie as a reachability problem for $e$), given the available initial resources for each path and model. 

This first simple modeling example conveys one of the basic intuitions behind our approach. Linear Logic can simultaneously model the control flow of our experiments (see the petri net token starting at linear proposition $e$ in Figure \ref{fig:RunningEg}) as well as all available resources and their consumption (see the petri net token at proposition $m$ in the same Figure). In fact this is achieved with a small number of logical operators ($\&,\otimes,\multimap,!$) reasoning over all execution paths and resources. As we will later see a more detailed approach requires us to describe this model in predicate rather than propositional Linear Logic, but this basic intuition will remain the same. More precisely our model can be extended with predicates for an execution stack in a way akin to abstract machines and transition systems, while logging the specific paths we are visiting and the resources we consume. 

\section{Linear Logic in Perspective}
We will now use the example we introduced in Figure \ref{fig:RunningEg}, to
provide a more detailed background on Linear Logic as it relates to our verification approach.
Linear Logic can be seen as a formal system
describing resource production, consumption and availability
\cite{scedrov1993brief,girard1995linear}. Contrary to classical or intuitionistic 
logic, Linear Logic explicitly models resources by disallowing structural 
rules of contraction and weakening \cite{di2006linear}. We briefly provide
some basic definitions, before proving the reachability
sequent: \colorbox{tableHeader}{$\Pi,M \vdash e$} that we saw earlier, 
where: $M=e,m$ represents the initial state of the tokens for our petri net (in Figure \ref{fig:RunningEg}) 
with $e$ modeling the control-flow and $m$ the resources of our model. Moreover, $\Pi=\pi_1,\pi_2,\pi_3,\pi_4$ represent our program transitions (modeled as linear logic formulas in the bottom right of Figure \ref{fig:RunningEg}). It thus follows that the sequent: \colorbox{tableHeader}{$\Pi,M \vdash e$}, asks if the training phase can successfully return, given the available initial resources for each path and each model. This corresponds to a petri-net reachability problem, with $e$ as the target. %

For the purposes of our application, a fruitful way to describe 
reasoning in linear logic is as
\textbf{a rewriting process operating over a \emph{multiset}}.
  In this setting, truths can be thought of 
  as \emph{transient resources} if they are currently available, 
  and can be consumed. 
The content of the multiset can be formally represented through a multiplicative
conjunction connective ($\otimes$), grouping the transient resources
together. For instance, in order to describe that both A and B hold, we can write $A \otimes B$. The bang operator $!$ is used to represent \emph{permanent
resources} stored in the multiset, that persist even after an instance of this 
resource has been consumed. \footnote{This is dual to the ($?$) operator 
that allows for permanent deletion} In essense, we use the bang operator to represent permanent 
and long-standing resources.

Resource transformations can be expressed using the lolli connector
($\multimap$) where the left-hand side of the linear implication describes
a subset of the multiset that is to be \textbf{replaced} by the right-hand side 
of the lolli ($\multimap$). Such implication rules are first-class (\ie they are themselves propositions
that can form larger propositions) and can be stored as either transient or permanent resources. For 
the purposes of this work, when we use the term \emph{instruction}, we will mean a linear implication
that is permanent.

For example given an initial configuration %
$A \otimes B \otimes C \otimes !(A \multimap D)$ we can derive a new
state by using the instruction $!(A \multimap D)$ that consumes
$A$ and replaces it with $D$. The resulting new memory state in this
case is: $D \otimes B \otimes C \otimes !(A \multimap D)$. In the
form of a sequent (\ie conditional assertion) the above derivation would be described as:
\[A \otimes B \otimes C \otimes !(A \multimap D) \vdash D \otimes B
  \otimes C \otimes !(A \multimap D)\] Finally, to represent resource
consumption alternatives, we can use the additive conjunction operator
($\&$). This operator can be thought of as a choice-operator, that is used to exhaustively explore the space of possible alternatives.

\subsection{Structuring proofs in Linear Logic}

    \begin{figure*}[h!t]
    \begin{prooftree}
    \AxiomC{$$}
    \RightLabel{\tiny{$id$}}
    \UnaryInfC{$e \vdash e$}
    \AxiomC{$$}
    \RightLabel{{$id$}}
    \UnaryInfC{$m \vdash m$}
    \AxiomC{$$}
    \RightLabel{{$id$}}
    \UnaryInfC{$t \vdash t$}
    \RightLabel{{$\otimes r$}}
    \BinaryInfC{$t,m \vdash t\otimes m$}
    \AxiomC{$$}
    \RightLabel{{$id$}}
    \UnaryInfC{$f_1 \vdash f_1$}
    \AxiomC{$$}
    \RightLabel{{$id$}}
    \UnaryInfC{$e \vdash e$}
    \RightLabel{{$\multimap l(\pi_3)$}}
    \BinaryInfC{$\pi_3,f_1 \vdash e$}
    \RightLabel{\colorbox{tableHeader}{{$\&l(f_2)$}}}
    \UnaryInfC{$\pi_3,f_1 \& f_2\vdash e$}
    \RightLabel{{$\multimap l(\pi_2)$}}
    \BinaryInfC{$\pi_2,\pi_3, t,m \vdash e$}
    \RightLabel{{$\multimap l(\pi_1)$}}
    \BinaryInfC{$\pi_1,\pi_2,\pi_3, e,m \vdash e$}
    \RightLabel{{$\pi_1,\pi_2, \pi_3 \in \Pi$}}
    \LeftLabel{$\Pi$}
    \UnaryInfC{$e,m \vdash e$}
    \end{prooftree}
    
    \caption{Proving that \colorbox{tableHeader}{$\Pi,M \vdash e$} (where $M=e,m$) for the $f_1$ path. Similarly for $f_2$, if we use \colorbox{tableHeader}{$\&l(f_1)$} instead of \colorbox{tableHeader}{$\&l(f_2)$}, in the highlighted rule} %
    \label{fig:petri-proof}
    
    \end{figure*}

In order to present the inference rules for the subset of Linear Logic (based on CLF \cite{watkins2003concurrentA}) we presented above, we give the following definitions:

\begin{itemize}
    \item Let $\Pi$ be a multiset of permanent linear implications. In our case $\Pi$ will be used to model a program in memory. Here the permanency of the implications naturally represents the permance of statements. We label such linear implications as $\pi_i$. From our running example from Figure \ref{fig:RunningEg} (bottom right corner) we have $\Pi = \pi_1 \otimes \pi_2 \otimes \pi_3 \otimes \pi_4$ where:\\
    \begin{itemize}
        \item $\pi_1$ = $!(e \multimap t)$
        \item $\pi_2$ = $!(t \otimes m \multimap f_1 \& f_2)$
        \item $\pi_3$ = $!(f_1 \multimap e)$
        \item $\pi_4$ = $!(f_2 \multimap e)$
    \end{itemize}
    \vspace{1em}
    As we previously explained, these correspond to the control-flow and resources of our experiment in the left-side of Figure \ref{fig:RunningEg}, expressed graphically with the petri-net in the top-right part of the same figure.
    From now on, we will using $\Pi$ to describe any similar multiset of program statements ($\pi_i$) expressed as linear implications.
    \item Let $M$ be a multiset of resources, modeling a program's memory state. From our running example in Figure \ref{fig:RunningEg} we have: 
    $M = e,m$.
    From now on, we will be using $M$ to describe the execution state of a program ($M_s$) as well as its initial resources ($M_p$) that are available at runtime. 
\end{itemize}

With these definitions, we can now describe a sequent calculus (in Figure \ref{fig:seqCalculus}) for our subset of Linear Logic, that allows us to rewrite (through inference) an initial state into subsequent states. The sequent calculus rules describe the introduction (in the left or the right hand side of a sequent) of three operators ($\otimes, \&, \multimap$) for multiplicative conjunction, additive conjunction and linear implication. The two additional rules 
for the right introduction of $\multimap$ and $\&$ are excluded from this figure given that they are not used in the examples we give. Finally, the last rule of Figure \ref{fig:seqCalculus} simply states the identity axiom for our system. Remember here that the multiset $\Pi$ is always preserved, since it represents permanent linear implications. Figure \ref{fig:petri-proof} uses this sequent calculus to prove that \colorbox{tableHeader}{$\Pi,M \vdash e$}.

The verbosity of proofs as the one we show in Figure \ref{fig:petri-proof}, becomes unmanageable (for human readers) in more realistic examples. %
Thus, we now show how these proofs can be simplified, by relaxing the requirements to (a) present the introduction of left and right $\otimes$ which can be implied by the sequent commas and (b) omitting the display of particular rules $\pi_i$ in the left side of implications, since they are permanently stored in $\Pi$:

\begin{prooftree}
\AxiomC{$e \vdash e$}
\RightLabel{$\multimap l(\pi_3)$}
\UnaryInfC{$f_1 \vdash e$}
\RightLabel{$\& l(f_2)$}
\UnaryInfC{$f_1 \& f_2 \vdash e$}
\RightLabel{$\multimap l(\pi_2)$}
\UnaryInfC{$t,m \vdash e$}
\RightLabel{$\multimap l(\pi_1)$}
\LeftLabel{$\Pi$}
\UnaryInfC{$e,m \vdash e$}
\end{prooftree}

The proof can be now read in a bottom-up fashion, where we start with the initial memory state $M=e,m$ and we subsequently apply $\pi_1$, $\pi_2$ and $\pi_3$ of $\Pi$ (making path choices like $\&l(f_2)$ as we move along) to obtain the final memory state: $M=e$. Simplifying even further, we can read and write the proof without focusing on the constant right-hand side of the sequent in each step. This is crucial for readability of longer formulas in predicate Linear Logic. Given that the right-hand side of the sequents remains constant, repeating them is not necessary. By minimizing the repeating right hand-side in our sequents we arrive at a proof style that is closer to a linear-logic transition or rewritting system:

\begin{prooftree}
    \AxiomC{$e\  \textcolor{gray}{\scriptscriptstyle\vdash e}$}
    \RightLabel{$\multimap l(\pi_3)$}
    \UnaryInfC{$f_1\  \textcolor{gray}{\scriptscriptstyle\vdash e}$}
    \RightLabel{$\& l(f_2)$}
    \UnaryInfC{$f_1 \& f_2\  \textcolor{gray}{\scriptscriptstyle\vdash e}$}
    \RightLabel{$\multimap l(\pi_2)$}
    \UnaryInfC{$t,m\  \textcolor{gray}{\scriptscriptstyle\vdash e}$}
    \RightLabel{$\multimap l(\pi_1)$}
    \LeftLabel{$\Pi$}
    \UnaryInfC{$e,m\  \textcolor{gray}{\scriptscriptstyle\vdash e}$}
\end{prooftree}

\section{Conclusion \& Future Work}

We have shown how and why Deep Learning experiments have critical requirements regarding data provenance and correctness of API calls that interact with hardware accelerators. These requirements can have a significant impact to the validity, efficiency and economics of Deep Learning experiments that nowdays permeate all areas of science and engineering. Indeed software mistakes relating to dataset slicing can contaminate experiments and lead to incorrect results. Moreover incorrect handling of APIs that interact with the hardware can lead to sub-optimal usage and untrustworthy conclusions. 

With these problems in mind we have investigated the use of Linear Logic for the analysis of Deep Learning experiments. We showed that primitives and operators of Linear Logic can be used to express: \textit{(i)} an abstract representation of the control flow of an experiment, \textit{(ii)} a set of available experimental resources, such as API calls to the underlying data-structures and hardware as well as \textit{(iii)} reasoning rules about the correct consumption of resources during experiments. We further noted that our proposed model is not only lightweight but also easy to comprehend having both a symbolic and a visual component. In terms of future work, we intend to expand upon our current presentation that relied on propositional Linear Logic, to include our full predicate model that covers monitoring of the execution stack and automatic tracing of our proofs. 

\bibliography{DeepLL} 

\end{document}